\documentclass[aps,superscriptaddress,groupedaddress]{revtex4}  
\usepackage{graphicx}  
\usepackage{dcolumn}   
\usepackage{bm}        
\usepackage{amssymb}   

\newcommand{\bdf}{$\beta$df}                               
\newcommand{\mcbdf}{mc-$\beta$df}                          

\begin{document}

\hspace{5.2in} \mbox{LA-UR-18-21034}

\title{$\beta$-delayed fission in $r$-process nucleosynthesis}

\author{M. R. Mumpower}
\affiliation{Theoretical Division, Los Alamos National Laboratory, Los Alamos, NM 87545, USA}

\author{T. Kawano}
\affiliation{Theoretical Division, Los Alamos National Laboratory, Los Alamos, NM 87545, USA}

\author{T. M. Sprouse}
\affiliation{Department of Physics, University of Notre Dame, Notre Dame, IN 46556, USA}

\author{N. Vassh}
\affiliation{Department of Physics, University of Notre Dame, Notre Dame, IN 46556, USA}

\author{E. M. Holmbeck}
\affiliation{Department of Physics, University of Notre Dame, Notre Dame, IN 46556, USA}

\author{R. Surman}
\affiliation{Department of Physics, University of Notre Dame, Notre Dame, IN 46556, USA}

\author{P. M{\"o}ller}
\affiliation{Theoretical Division, Los Alamos National Laboratory, Los Alamos, NM 87545, USA}

\date{\today}

\begin{abstract}

We present $\beta$-delayed neutron emission and $\beta$-delayed fission (\bdf) calculations for heavy, neutron-rich nuclei using the coupled Quasi-Particle Random Phase Approximation plus Hauser-Feshbach (QRPA+HF) approach.
From the initial population of a compound nucleus after $\beta$-decay, we follow the statistical decay taking into account competition between neutrons, $\gamma$-rays, and fission.
We find a region of the chart of nuclides where the probability of \bdf\ is $\sim100$\%, that likely prevents the production of superheavy elements in nature.
For a subset of nuclei near the neutron dripline, neutron multiplicity and the probability of fission are both large, leading to the intriguing possibility of multi-chance \bdf, a new decay mode for extremely neutron-rich heavy nuclei.
In this new decay mode, $\beta$-decay can be followed by multiple neutron emission leading to subsequent daughter generations which each have a probability to fission.
We explore the impact of \bdf\ in rapid neutron capture process ($r$-process) nucleosynthesis in the tidal ejecta of a neutron star--neutron star merger
and show that it is a key fission channel that shapes the final abundances near the second $r$-process peak.

\end{abstract}

\maketitle

One of the greatest challenges in theoretical physics is the determination of the astrophysical location where the heaviest elements on the periodic table are synthesized \cite{Burbidge+57, Cameron+57}. 
Solving this problem couples together the quantum theory of nuclei with the description of cataclysmic astrophysical events such as supernovae or merging compact objects \cite{Arnould+07}. 
Recent gravitational wave and electromagnetic observations suggest that the rapid neutron capture process ($r$ process) of nucleosynthesis occurs in the merger of two neutron stars (NSM) \cite{AbbottGW170817, Cowperthwaite+17}. 
The $r$ process in NSMs can create heavy, neutron-rich elements through a series of neutron captures and $\beta$-decays among other nuclear reactions \cite{Mumpower+16r}.
Many open questions remain regarding the type of conditions that can occur in this environment, but the bulk of the composition of the ejecta is thought to be determined by two primary types of conditions \cite{Kajino+17}. 
The dynamical---or tidally ejected---material flung off at relatively high velocities is expected to be very neutron-rich, capable of producing the actinides that undergo fission \cite{Lattimer+74, Meyer+89, Freiburghaus+99}. 
The other type of environment that may be present is a viscous and/or neutrino-driven wind component, which is closer to the center of the merging interface and is believed to have a range of neutron-richness \cite{Surman+08, Wanajo+14, Just+15, Rosswog+17, Siegel+17}. 
Observations of the GW170817 transient are consistent with the radioactive afterglow of high-opacity, lanthanide-rich material \cite{Tanvir+17}, which intriguingly seems to require some hundredths of a solar mass \cite{Kasen+17} of highly neutron-rich ejecta, characteristic of dynamical ejecta conditions. 

The study of low-electron-fraction (neutron-rich) ejecta is important as it opens the possibility to gain insight into unmeasured reactions, decays, and fission that transmute the heaviest exotic nuclei \cite{Mumpower+17}. 
Fission in the $r$ process is of particular interest; simulations show that an $r$ process that undergoes multiple fission cycles---where the $r$-process path terminates via fission and fission products continue to capture neutrons---results in a consistent pattern of abundances between the second and third $r$-process peaks for a given set of nuclear physics inputs \cite{Beun+08, Temis+15}. 
This potentially offers an explanation for the consistent $52\leq Z \leq 82$ $r$-process abundance patterns observed in a wide range of stars throughout our galaxy that match the solar $r$-process pattern \cite{Sneden+08}. 
The final abundance pattern of a fission recycling $r$ process is largely shaped by fission properties \cite{Cote+17}: neutron-induced fission rates \cite{Panov+03, Pinedo+07}, spontaneous fission rates \cite{Panov+13}, $\beta$-delayed fission probabilities \cite{Thielemann+83, Ghys+15}, neutrino-induced fission rates \cite{Qian+02, Kolbe+04}, and fission fragment yields \cite{Eichler+14, Temis+15}. 
Here we perform a fresh investigation of $\beta$-delayed fission (\bdf) and examine its role in a fission-recycling $r$ process. 

The process of $\beta$-delayed fission (\bdf) is a two-step nuclear decay process that couples $\beta$-decay and fission \cite{Andreyev+13}. 
In \bdf, a precursor nucleus ($Z$,$A$) with $Z$ protons and $A$ nucleons,  $\beta^{\pm}$-decays into a daughter nucleus ($Z\mp1$,$A$) that has a probability to fission. 
The study of this low-energy decay mode remains a great challenge experimentally due to the rare branching ratios relative to $\alpha$-decay \cite{Elseviers+13, Liberati+13, Truesdale+16}. 
For nuclei that may participate in the $r$ process, \bdf\ may be the dominant branching mode due to low fission barriers \cite{Moller+15}. 
To model the complicated processes after $\beta$-decay, we have recently shown that the competition between neutron and $\gamma$-ray emission should be included \cite{Mumpower+16b}. 
This theoretical interpretation is in agreement with recent experimental results for light neutron-rich nuclei \cite{Spyrou+16}. 

\begin{figure*}
 \begin{center}
  \centerline{\includegraphics[width=150mm]{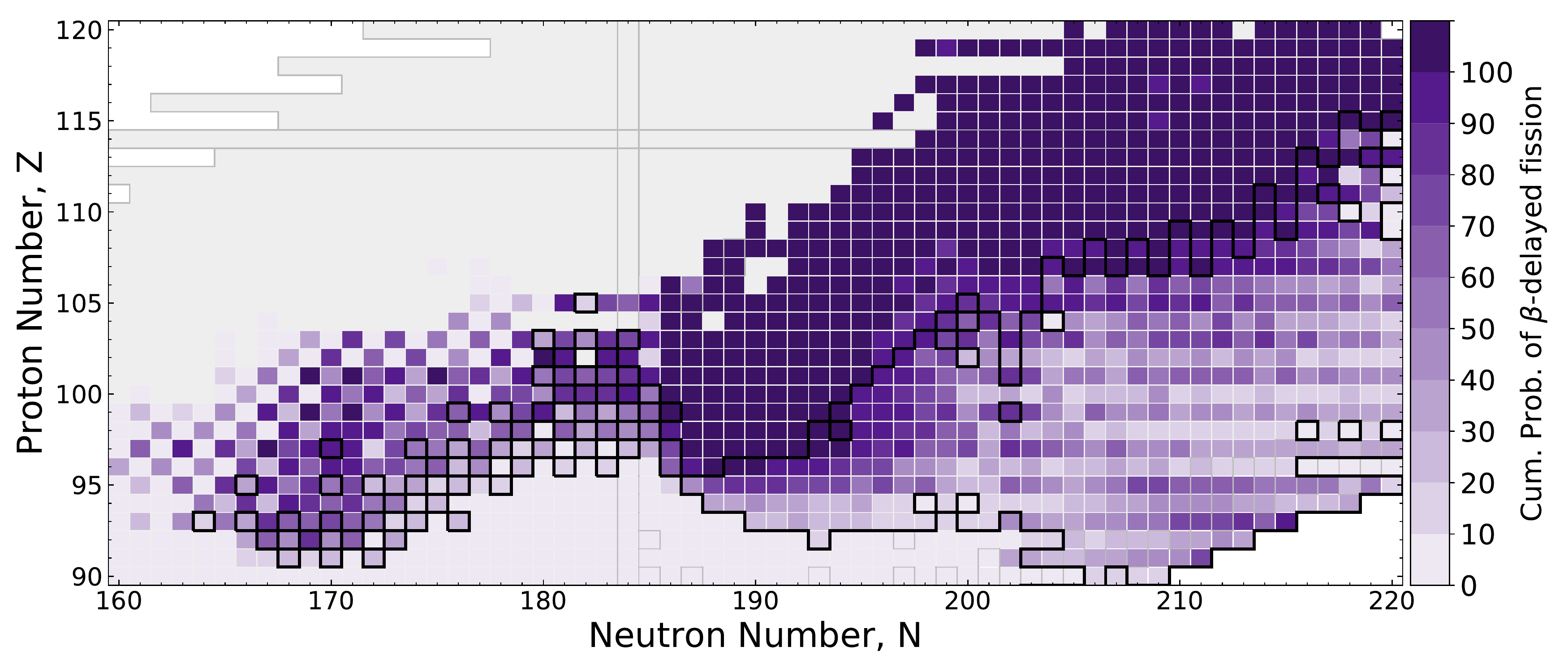}}
  \caption{\label{fig:results} (Color online) Cumulative probability ($P_\textrm{\small f}$) of $\beta$-delayed fission (\bdf) for neutron-rich nuclei using the QRPA+HF framework. The nuclei which exhibit multi-chance \bdf\ are outlined in a black bounding box satisfying the constraint: $\sum_{j=1}^{10}P_{j\textrm{\small f}} \geqslant 10\%$. Grey shading denotes nuclei that are bound by proton or neutron emission in FRDM2012 and have $P_\textrm{\small f}=0$.}
 \end{center}
\end{figure*}

In this Letter, we extend our coupled Quasi-Particle Random Phase Approximation and Hauser-Feshbach (QRPA+HF) framework to describe nuclei that may undergo \bdf. 
Our treatment leads us to identify a region of the chart of nuclides where the probability of \bdf\ is near 100\%. 
This region lies between the neutron dripline and the predicted superheavy island of stability, blocking the decay of $r$-process species into possibly stable superheavy elements. 
We discuss a new decay mode for neutron-rich, heavy, $r$-process nuclei: multi-chance $\beta$-delayed fission (\mcbdf) in which $\beta$-decay may be followed by fission in daughter generations after each stage of neutron emission, similar to multi-chance neutron-induced fission. 
We show that \bdf\ impacts the final abundances of the $r$-process since it operates on the timescale of $\beta$-decay, thus influencing where fission fragments are distributed during the end of fission recycling. 

The QRPA+HF framework is discussed in Ref.~\cite{Mumpower+16b} in the context of $\beta$-delayed neutron emission. 
We provide below a brief overview of the model and the necessary enhancement when describing fission. 
We use the Los Alamos BeoH code, now at version 3.5.0, to follow the statistical de-excitation after $\beta$-decay. 
This code begins with the initial population of a compound nucleus, following Gamow-Teller $\beta$-decay treated in the QRPA \cite{Moller+97} built on top of the 2012 version of the Finite-Range Droplet Model (FRDM2012) \cite{FRDM2012a, FRDM2012b}. 
The subsequent statistical decay of this nucleus is followed by including the competition between neutron emission and $\gamma$ de-excitation, producing output of particle spectra and branching ratios. 
For the neutron-rich nuclei studied in this paper, standard optical model and $\gamma$-strength function choices are used along with a Glibert-Cameron level density with shell corrections included following the prescription of Ignatyuk \cite{Mumpower+16b}. 

When describing nuclei that may fission, an additional transmission coefficient must be calculated. 
We assume that this transmission coefficient takes the Hill-Wheeler functional form \cite{Hill+53} representing transmission through a single-hump parabolic barrier dependent on predictions of barrier heights \cite{Moller+15} with curvature parameter from Ref.~\cite{Thielemann+83}. 
We note the presence of additional shorter and wider fission barriers does not have a qualitative impact on our conclusions. 
The level density at the fission saddle point enters into the total fission transmission coefficient in a multiplicative manner, playing a significant role in determining whether or not fission occurs in each daughter generation. 
The fission level density is larger in deformed configurations than in the ground state due to additional shape degrees of freedom and other collective effects. 
Levels on top of the barrier are built with a rotational enhancement factor which leads to an increase of the level density by roughly a factor of four relative to the ground state \cite{Bjornholm+73, Iljinov+92}. 

The statistical de-excitation process is followed in each subsequent daughter generation until all of the initial excitation energy is exhausted. 
It is therefore critical that the decay chain have consistently calculated nuclear properties from the QRPA strength to neutron separation energies and fission barrier heights. 
We have found that a good upper bound to the number of daughter generations is $10$ neutrons away from the first daughter nucleus for $r$-process nuclei \cite{Mumpower+16b}. 
The statistical decay must end in either the population of a daughter generation's ground state or in fission, thus the summation of the probabilities to emit a neutron or to fission must be equal to unity:
\begin{equation}
 1 = \sum_{j=0}^{10}(P_{j\textrm{\small n}} + P_{j\textrm{\small f}}) = P_{\textrm{\small n}} + P_{\textrm{\small f}} \quad ,
 \label{eqn:cP}
\end{equation}
where $P_{j\textrm{\small n}}$ is the probability to emit $j$ neutrons, $P_{j\textrm{\small f}}$ is the probability for the $j$-th compound nucleus to fission, and the cumulative values of these two quantities are denoted $P_\textrm{\small n}$ and $P_\textrm{\small f}$, respectively. 
Regular \bdf\ is defined by $P_{0\textrm{\small f}}\neq0$ while multi-chance $\beta$-delayed fission occurs when $P_{j\textrm{\small f}}\neq0$ for $j>0$, which we describe shortly. 

We calculate $\beta$-delayed neutron emission and fission probabilities using the QRPA+HF framework for all neutron-rich nuclei from stability to extreme neutron excess. 
The cumulative probability for \bdf, $P_{\textrm{\small f}}$, is shown in Fig.~\ref{fig:results} for a subset of these nuclei relevant to the $r$ process.

There are two key features present in Fig.~\ref{fig:results} that may have important consequences for $r$-process nucleosynthesis. 
The first is the large subset of nuclei beyond the predicted $N=184$ shell closure in FRDM2012 that have a $P_{\textrm{\small f}}=100$\%, i.e., the $\beta$-decay chain always ends in fission rather than the population of the ground state of any $\beta$ daughter generation, $P_\textrm{\small n}=0$. 
In this region, the nuclear flow of the $r$ process can no longer increase in proton number via $\beta$-decay, which can result in termination via \bdf. 
This idea was first explored in Ref.~\cite{Thielemann+83}, and we find the \bdf\ region in this work extends much further in the $NZ$-plane and has relatively lower probabilities for nuclei near ($Z$,$N$)$\sim$($95$,$170$). 
This extended region of high \bdf\ probability may also prevent the production of superheavy elements (near the crossing of the $Z=114$ and $N=184$ closed shells in Fig.~\ref{fig:results}) in nature, by blocking decay pathways between the $r$-process path and the island of stability. 
The thwarting of superheavy element production by regions of high \bdf\ probability was also examined in Ref.~\cite{Thielemann+83}, however only nuclei with proton number less than $Z=100$ (lower than the predicted island of stability) were considered. 
Older \cite{Meldner+72} as well as more recent calculations \cite{Petermann+12} proposed a neutron-rich pathway where nuclei may circumvent fission and subsequently populate superheavy nuclei by chains of $\beta$- and $\alpha$- decays. 
The extended region of high \bdf\ probability in the present work makes this possibility more difficult for the nuclear flow to achieve, a point we return to in the discussion of our nucleosynthesis calculations below. 
Figure \ref{fig:results} shows that closer to the neutron dripline the cumulative probability for \bdf\ decreases. 
Many factors can influence whether or not the nuclear flow can continue higher at extreme neutron excess, including how close the low-probability region comes to the uncertain location of the neutron dripline \cite{Erler+12}. 

Very close to the neutron dripline an interesting phenomenon arises. 
Here nuclei tend to have very large $\beta$-decay Q-values and large neutron multiplicities (on the order of 3-10) in the decay chains towards stability due to small neutron separation energies \cite{Mumpower+16b}. 
Since the nuclei along these decay chains have relatively low fission barriers (as predicted with different nuclear potentials \cite{Howard+80, Jachimowicz+17}), it opens the possibility for each of the populated daughter generation to fission after $\beta$-decay, $P_{j\textrm{\small f}}\neq0$ for some $j$. 
We call this phenomenon multi-chance $\beta$-delayed fission (\mcbdf) analogous to multi-chance neutron-induced fission. 
This phenomenon can occur in decay chains where neutron multiplicity is greater than zero and there is sufficient cumulative probability to fission. 
We define this constraint to be $\sum_{j=1}^{10}P_{j\textrm{\small f}} \geqslant 10\%$ with the $j=0$ component or regular \bdf\ left out of the summation. 
The nuclei which exhibit \mcbdf\ are a subset of the most neutron-rich nuclei that are predicted to undergo \bdf, occupying roughly half of the \bdf\ ``real estate" towards the neutron-dripline. 

The existence of \mcbdf\ is the second key feature of Fig.~\ref{fig:results} that can impact the $r$ process, via changes to how the nuclear flow proceeds from heavy to light nuclei during fission recycling. 
Rather than only considering the daughter ($j=0$) that fissions, as in \bdf, each possible daughter generation ($j>0$) may fission. 
Thus the population of light nuclei during the $r$ process from this fission channel is actually a superposition of fission fragment yields of a chain of all the daughter nuclei. 
Another consequence of this new decay mode is that it may increase or decrease the auxiliary source of free neutrons during the decay back to stability known as freeze-out \cite{Mumpower+15}. 
An increase or decrease in available neutrons depends on the nuclear mass surface and the number of prompt neutrons after scission, among other things. 

To gauge the impact of the \bdf\ properties, we perform $r$-process nuclear network calculations using version $1.0$ of PRISM (Portable Routines for Integrated nucleoSynthesis Modeling) \cite{Mumpower+17, Cote+17, Sprouse+18}. 
This reaction network is written in fully object-oriented Fortran supporting unique input control of all reaction channels. 
The baseline nuclear model used for our calculations is FRDM2012. 
Standard $r$-process reaction channels, i.e., neutron capture, photodissociation, $\beta$-decay, $\beta$-delayed neutron emission, alpha decay, and neutron-induced fission are included along with the \bdf\ rates and branching ratios from this work. 
The reaction rates of $r$-process nuclei are calculated using the LANL statistical Hauser-Feshbach code CoH \cite{Kawano+16}, also at version 3.5.0. 
Fission fragment distributions from Kodama et al.~\cite{Kodama+75} are used for neutron-induced, beta-delayed, and spontaneous fission yields unless stated otherwise. 
Evaluated data from NUBASE2016 \cite{NUBASE2016}, including measured spontaneous fission rates, and measured masses from the AME2012 \cite{AME2012} are used when available. For astrophysical conditions, we choose neutron star merger trajectories corresponding to the ``slow'' neutron-rich dynamical ejecta of Ref.~\cite{Temis+15} (with nuclear reheating included) as well as the low entropy ($s\sim10$) with very low initial electron fraction ($Y_e\sim0.01$) dynamical ejecta from Ref.~\cite{Goriely+11}. 

\begin{figure}
 \begin{center}
  \centerline{\includegraphics[width=\columnwidth]{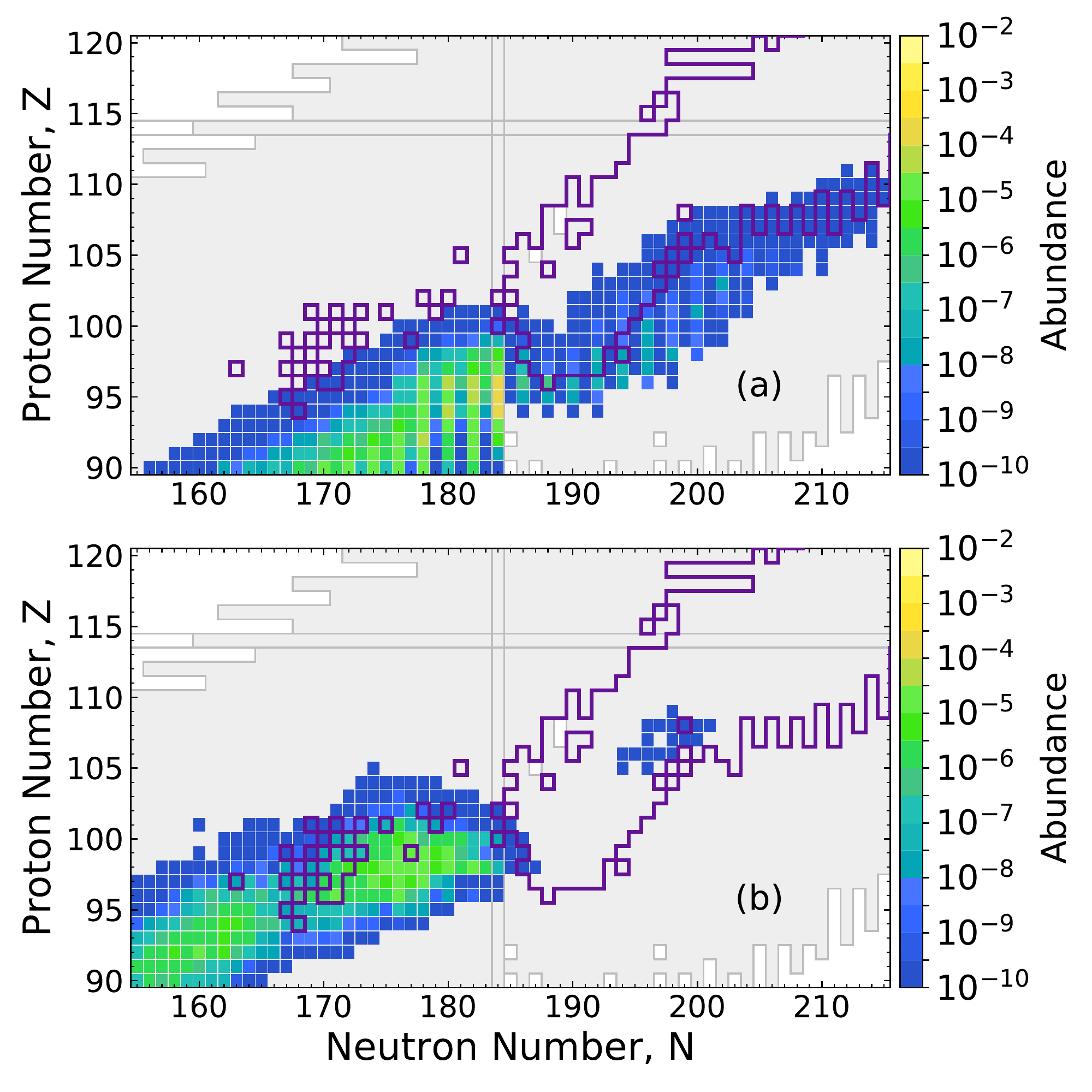}}
  \caption{\label{fig:supers} (Color online) The $r$-process abundances at a time (a) just before the nuclear flow encounters the $P_{\textrm{f}}\geqslant90\%$ region from Fig.~\ref{fig:results} (outlined in purple) and (b) after encountering this high probability \bdf\ region. The network calculations shown here include only \bdf\ and experimental spontaneous fission channels and are for astrophysical conditions from \cite{Temis+15}.}
 \end{center}
\end{figure}

First we consider the possible production of superheavy elements in our $r$-process calculations. 
For all example calculations considered we find that neutron-induced fission rates calculated with FRDM2012 masses and barrier heights can alone prevent the formation of superheavy elements, in agreement with \cite{Boleu+72, Howard+74, Petermann+12}. 
In such calculations, neutron-induced fission terminates the $r$ process around $A\approx 290$, lower than the crossing of the shell closures in Fig.~\ref{fig:results}. 
However, since prescriptions for astrophysical capture rates can vary by several orders of magnitude, it is important to consider the effect of \bdf\ alone (e.g. in the absence of neutron-induced fission). 
We repeat our network calculations with only the \bdf\ and experimental spontaneous fission \cite{NUBASE2016} channels included and show the evolution of $r$-process abundances in Fig.~\ref{fig:supers}. 
Figure \ref{fig:supers} demonstrates that $A\gtrsim 290$ nuclei are (a) populated at a time before the nuclear flow encounters the region from Fig.~\ref{fig:results} with a \bdf\ probability $\geqslant90\%$ and (b) not populated after moving through this high probability \bdf\ region. 
Therefore, even in the absence of neutron-induced fission, \bdf\ is itself sufficient to prevent the production of superheavy nuclei by blocking decay pathways to stability. 

\begin{figure}
 \begin{center}
  \centerline{\includegraphics[width=\columnwidth]{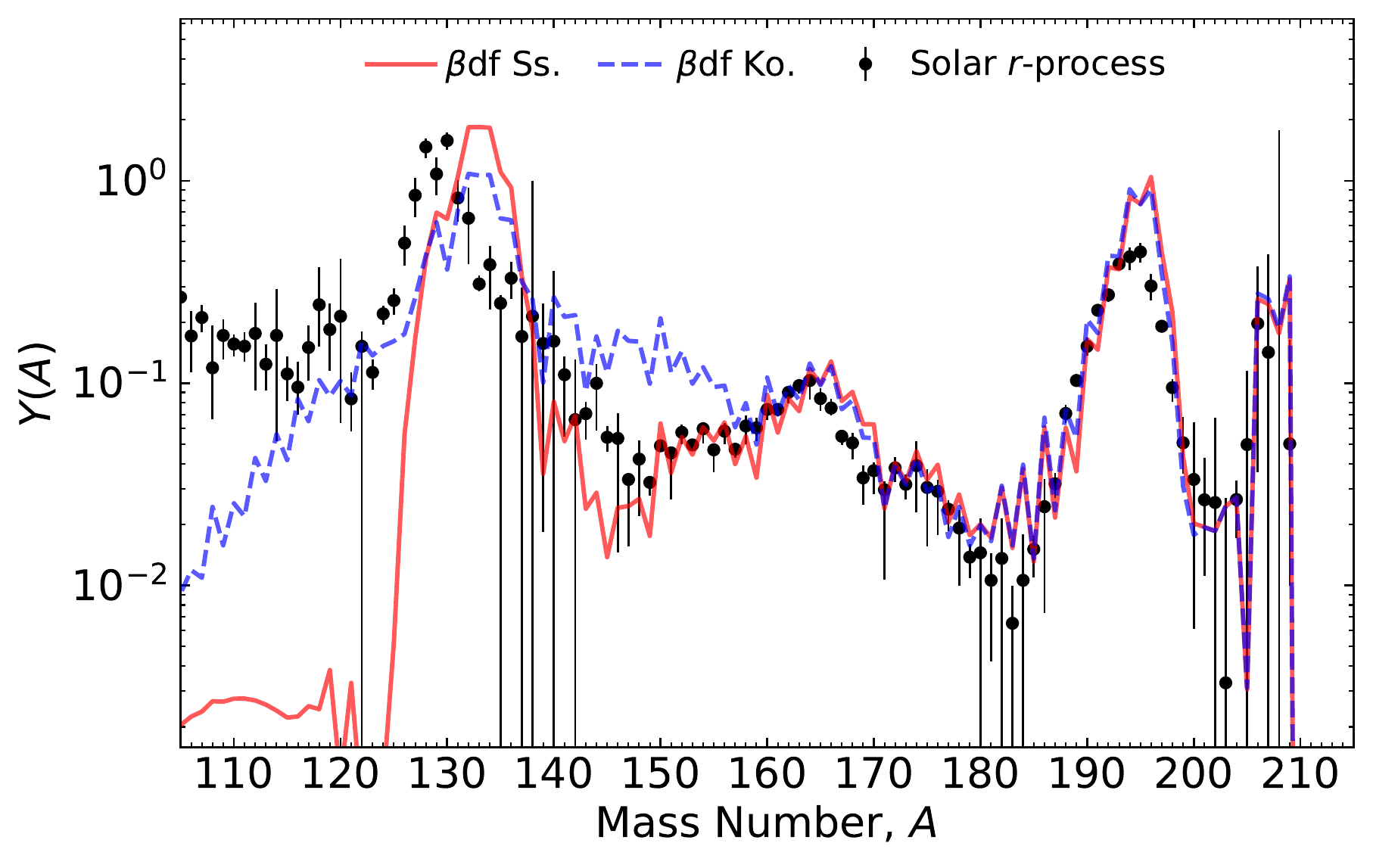}}
  \caption{\label{fig:ab} (Color online) Final abundances of a neutron-star merger trajectory \cite{Goriely+11} at $1$Gyr using 50/50 symmetric fission yields (\bdf~Ss., red) as compared to the yield distribution of Kodama et al. (\bdf~Ko., blue, dashed) for the \bdf\ channel. Solar data from Ref.~\cite{Sneden+08} shown in black dots. }
 \end{center}
\end{figure}

As indicated above, our calculations show that when all fission channels are included the $r$ process terminates via neutron-induced fission. 
It might be expected, then, that neutron-induced fission is largely responsible for shaping the final $r$-process abundance pattern.  
To show an example of how \bdf\ can influence the $r$-process pattern, we apply two different fission fragment distribution schemes. 
In the first, we replace the Kodama fission fragment distribution of the neutron-induced fission yields with a simple, symmetric split which assigns an atomic mass number half that of the fissioning system to each of the fragments (and no neutrons are emitted). 
The \bdf\ fragments retain the Kodama yields (\bdf~Ko.). For the second fission scheme, we use simple symmetric splits for both neutron-induced fission and \bdf\ channels (\bdf~Ss.). 
When applied to all fissioning nuclei, a simple symmetric split produces an abundance pattern in which the fission products are deposited directly in the $A\sim 130$ region, leading to a sharp, well-defined second $r$-process peak as shown by the solid line in Fig.~\ref{fig:ab}. 
The Kodama yields, on the other hand, use a double Gaussian distribution, placing fission products in a wide range of nuclei. 
This distribution results in a weaker, broader second peak that extends over a range of atomic mass number. 
If neutron-induced fission dominated at all times, the effect of changing the \bdf\ fragment distribution would be small and the abundance patterns with the two fission schemes should be similar. 
Instead, Fig.~\ref{fig:ab} shows the final patterns are distinct. 
The clear signature of the Kodama yields appears with the \bdf~Ko.\ scheme, indicating the \bdf\ channel plays a key role in shaping the $A\sim130$ peak. 

\begin{figure}
 \begin{center}
  \centerline{\includegraphics[width=\columnwidth]{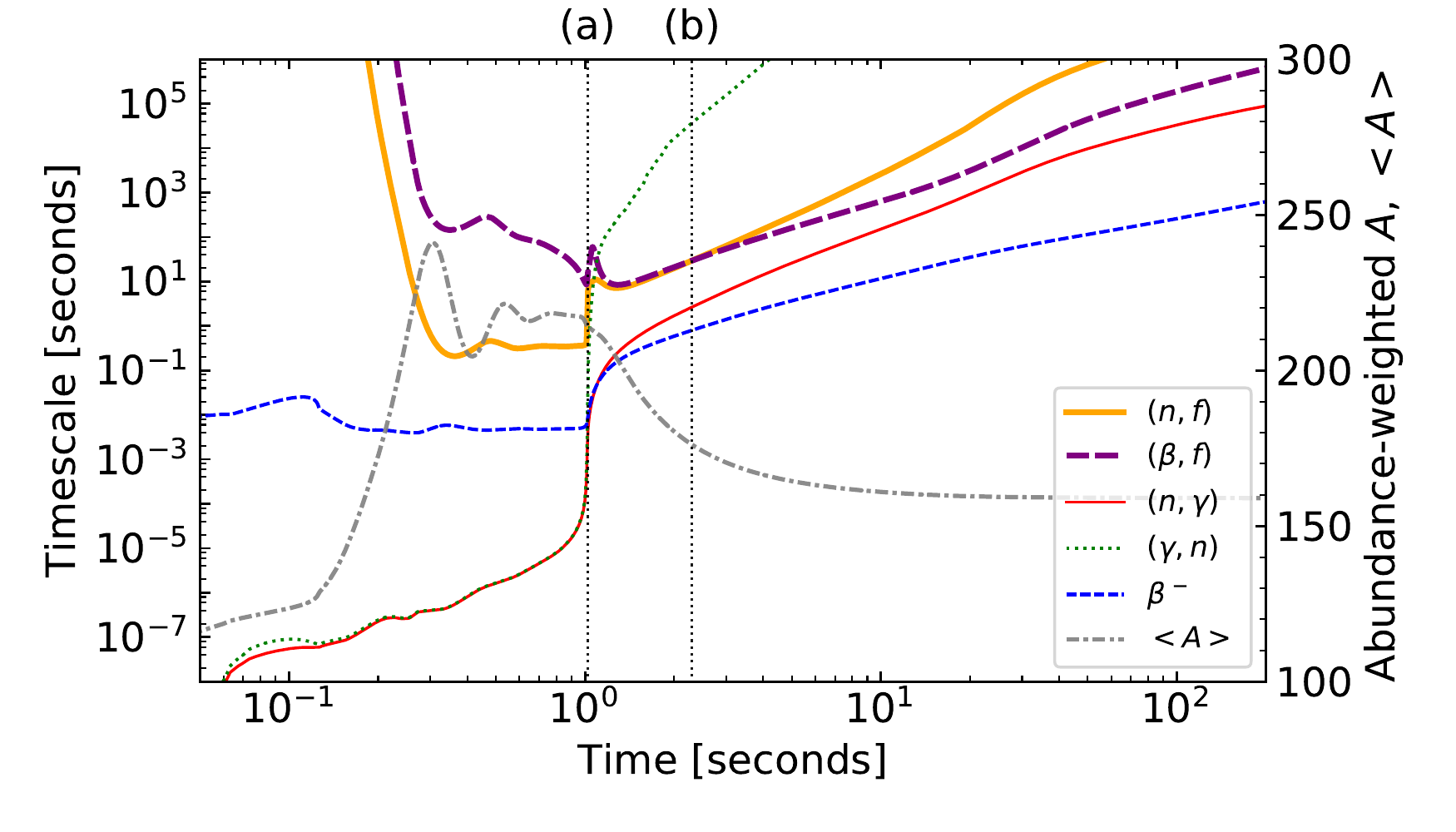}}
  \caption{\label{fig:tau} (Color online) A selection of abundance-weighted timescales versus time for the $r$-process phase of a neutron-star merger event \cite{Goriely+11}. The vertical line (a) indicates the time at which neutron capture (n,$\gamma$) (thin solid line) falls out of equilibrium with photodissociation ($\gamma$,n) (thin dotted line) and begins to compete with $\beta$-decay (thin dashed line), while vertical line (b) demarks the point when the \bdf\ ($\beta$,f) (thick dashed line) timescale is faster than the timescale for neutron-induced fission (n,f) (thick solid line). The abundance-weighted average mass number is shown by the dot-dashed line, read from the right Y-axis.}
 \end{center}
\end{figure}

The precise mechanism of this influence can be understood from an investigation of the $r$-process dynamics. 
Figure \ref{fig:tau} shows the relative importance of select nuclear reaction channels as the $r$ process proceeds in neutron-rich conditions. 
These abundance-weighted timescales indicate an overall global perspective of the dynamics during the $r$ process \cite{Mumpower+12a}. 
The early $r$ process is characterized by (n,$\gamma$)-($\gamma$,n) equilibrium. 
Fission becomes an important channel once the nuclear flow moves beyond the $N=184$ closed neutron shell, around several hundred milliseconds. 
The onset of fission recycling is indicated by the peak in the abundance-weighted average $A$, shown by the dot-dashed line in Fig.~\ref{fig:tau}. 
During these early times, neutron-induced fission dominates fission recycling, and it continues to be the primary fission channel throughout most of the fission recycling phase for these astrophysical conditions. 
As (n,$\gamma$)-($\gamma$,n) fails, indicated in Fig.~\ref{fig:tau} by the left vertical dashed line (a) at $t\sim 1$~s, a competition between neutron-capture and $\beta$-decay arises. 
At this stage, the neutron abundance is dropping rapidly and thus fission fragments cannot change significantly in $A$ via neutron captures as they decay back to stability. 
At about $t\sim 2.3$~s, the right vertical dashed line labeled (b) in Fig.~\ref{fig:tau}, \bdf\ becomes the primary fission channel while nuclei are decaying. 
Thus the late-stage or final fission cycle, which is increasingly dominated by \bdf, will dictate the final form of the abundances of lighter-mass nuclei near the $A=130$ peak, as supported by Fig.~\ref{fig:ab}. 

\begin{figure}
 \begin{center}
  \centerline{\includegraphics[width=\columnwidth]{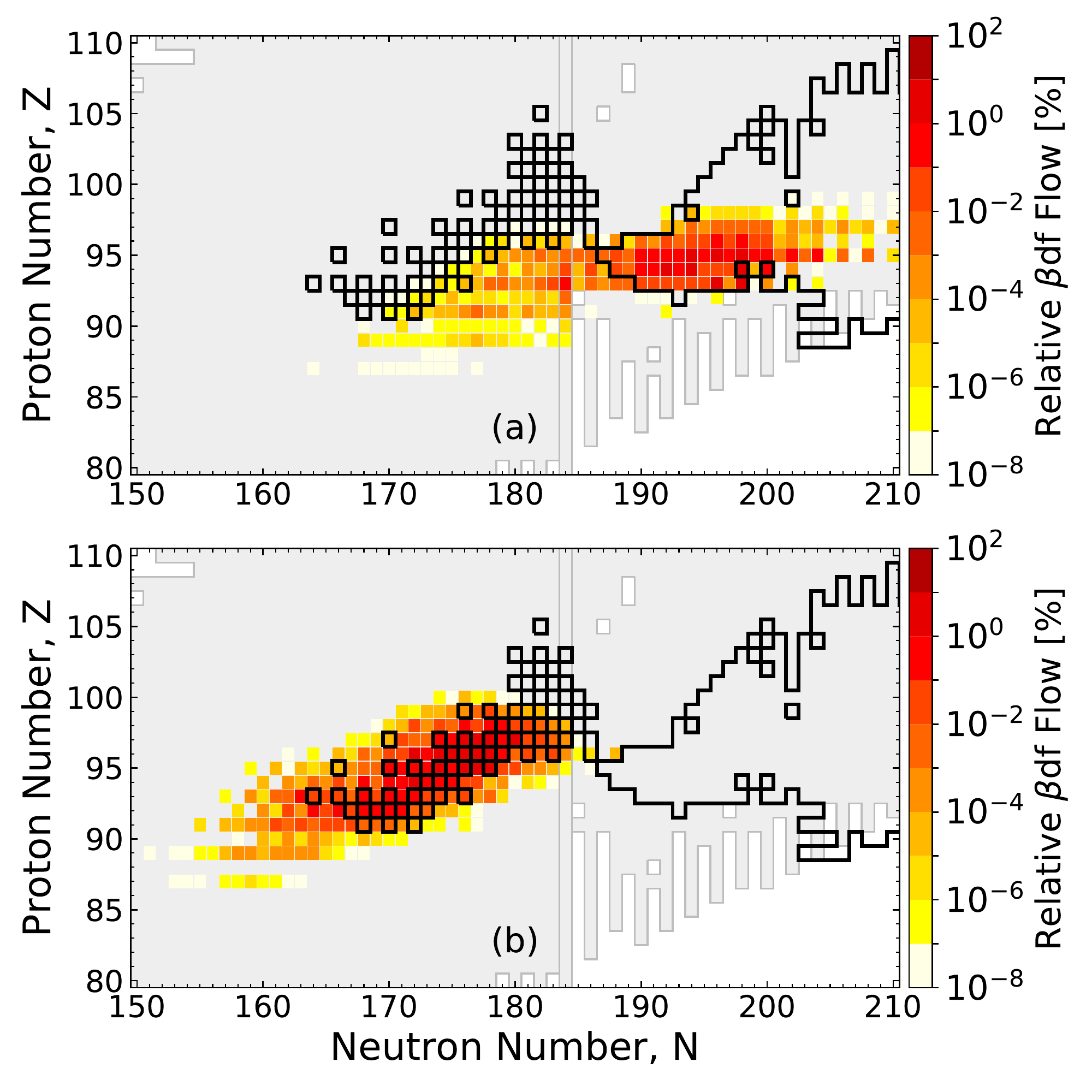}}
  \caption{\label{fig:flows} (Color online) Relative \bdf\ flow as compared to the total fission flow for all nuclei at an early time (a) when neutron-induced fission dominates and at a later time (b) when \bdf\ becomes dominant. Selected times correspond to the dotted vertical lines in Fig.~\ref{fig:tau}. At both times, \mcbdf\ exhibits large flow indicated by the black outlined region as in Fig.~\ref{fig:results}.}
 \end{center}
\end{figure}

We now show that \mcbdf\ is an integral part of the \bdf\ channel. 
To understand the dynamics in the region of fissioning nuclei, it is informative to look at the relative fission flows. 
Reaction flows, $F=Y(Z,N)\times\lambda$, where $Y(Z,N)$ is the abundance and $\lambda$ is the rate in $s^{-1}$, permit an observation of the hotspots for particular reaction channels in the $NZ$-plane. 
In Fig.~\ref{fig:flows} we examine the \bdf\ flows, denoted $F(Z,N)_{\beta,\textrm{\small f}}$, at an early and late time in the $r$-process evolution.
The colors indicate the relative \bdf\ flow, $F(Z,N)_{\beta,\textrm{\small f}}/\dot{Y}_\textrm{\small f}$, where $\dot{Y}_\textrm{\small f} =\sum_{Z,N}\left[F(Z,N)_{\beta,\textrm{\small f}}+F(Z,N)_{\textrm{\small n},\textrm{\small f}}\right]$ is the total fission flow of all nuclei in the network at a given time with $F(Z,N)_{\textrm{\small n},\textrm{\small f}}$ representing the neutron-induced fission flow. 
We neglect spontaneous fission in this sum since it is not expected to substantially contribute far from stability using parameterized models, e.g., from Ref.~\cite{Zagrebaev+11,Petermann+12}, on the timescale of interest for neutron-induced or \bdf. 
The region where the cumulative $\beta$-delayed multi-chance fission probability exceeds $10\%$ (as described in Fig.~\ref{fig:results}) is outlined in Fig.~\ref{fig:flows} in black. 
Nuclei exhibiting \mcbdf\ are active both at early times when neutron-induced fission contributes significantly (a), and at later times when the neutron flux has been exhausted and the \bdf\ channel begins to dominate (b). 
Since the bulk of the high-mass flow goes through the \mcbdf\ region, \mcbdf\ contributes significantly to the \bdf\ flow until very late times in the simulation. 
It is only once the $r$-process path has drawn closer to stability that regular, first-chance \bdf\ takes over. 

In summary we have extended the QRPA+HF framework to describe $\beta$-delayed fission (\bdf) for nuclei that may participate in neutron-rich heavy element nucleosynthesis. 
We isolated a region of nuclear chart with \bdf\ probability near 100\% that has an extended range in proton number. 
We have demonstrated that this extended \bdf\ region can prevent the formation of the superheavy elements in nature by the rapid neutron-capture process. 
We have identified a new decay mode, multi-chance $\beta$-delayed fission (\mcbdf), which operates near the neutron dripline. 
We have shown that \mcbdf\ contributes to the bulk of the \bdf\ flow in the dynamical ejecta of a neutron star merger $r$ process and therefore can impact the final abundances near the second $r$-process peak. 
Since the predicted large region of \bdf\ operates on the timescale of $\beta$-decay, it may also impact the calculation of the kilonova afterglow. 
Thus, future efforts to understand \bdf\ and \mcbdf\ will be crucial to unlocking the mysteries surrounding the formation of heavy elements and the interpretation of electromagnetic transients of nucleosynthetic events. 

The authors would like to thank Oleg Korobkin and Chris Fryer for useful discussions. 
This work was supported in part by the FIRE: Fission In $R$-process Elements collaboration (MM, TK, NV, and RS). 
MM, TK and PM were supported by the National Nuclear Security Administration of the U.S. Department of Energy at Los Alamos National Laboratory under Contract No. DE-AC52-06NA25396. 
TMS, EMH, and RS were supported by U.S. Department of Energy contract No. DE-SC0013039. 

\bibliographystyle{unsrt}
\bibliography{refs}

\begin{thebibliography}{10}

\bibitem{Burbidge+57}
E.~M. {Burbidge}, G.~R. {Burbidge}, W.~A. {Fowler}, and F.~{Hoyle}.
\newblock {Synthesis of the Elements in Stars}.
\newblock {\em Reviews of Modern Physics}, 29:547--650, 1957.

\bibitem{Cameron+57}
A.~G.~W. {Cameron}.
\newblock {Nuclear Reactions in Stars and Nucleogenesis}.
\newblock {\em Chalk River Reports}, 1957.

\bibitem{Arnould+07}
M.~{Arnould}, S.~{Goriely}, and K.~{Takahashi}.
\newblock {The r-process of stellar nucleosynthesis: Astrophysics and nuclear
  physics achievements and mysteries}.
\newblock {\em Physics Reports}, 450:97--213, September 2007.

\bibitem{AbbottGW170817}
B.~P. {Abbott \textit{et al.}}
\newblock Gw170817: Observation of gravitational waves from a binary neutron
  star inspiral.
\newblock {\em Phys.\ Rev.\ Lett.}, 119:161101, Oct 2017.

\bibitem{Cowperthwaite+17}
P.~S. {Cowperthwaite \textit{et al.}}
\newblock The electromagnetic counterpart of the binary neutron star merger
  ligo/virgo gw170817. ii. uv, optical, and near-infrared light curves and
  comparison to kilonova models.
\newblock {\em The Astrophysical Journal Letters}, 848(2):L17, 2017.

\bibitem{Mumpower+16r}
M.~R. {Mumpower}, R.~{Surman}, G.~C. {McLaughlin}, and A.~{Aprahamian}.
\newblock {\em Progress in Particle and Nuclear Physics}, 86:86--126, January
  2016.

\bibitem{Kajino+17}
Toshitaka Kajino and Grant~J Mathews.
\newblock Impact of new data for neutron-rich heavy nuclei on theoretical
  models for r -process nucleosynthesis.
\newblock {\em Reports on Progress in Physics}, 80(8):084901, 2017.

\bibitem{Lattimer+74}
J.~M. {Lattimer} and D.~N. {Schramm}.
\newblock {Black-hole-neutron-star collisions}.
\newblock {\em Astrophys.\ J.}, 192:L145--L147, September 1974.

\bibitem{Meyer+89}
B.~S. {Meyer}.
\newblock {Decompression of initially cold neutron star matter - A mechanism
  for the r-process?}
\newblock {\em Astrophys.\ J.}, 343:254--276, August 1989.

\bibitem{Freiburghaus+99}
C.~{Freiburghaus}, S.~{Rosswog}, and F.-K. {Thielemann}.
\newblock {R-Process in Neutron Star Mergers}.
\newblock {\em Astrophys.\ J.}, 525:L121--L124, November 1999.

\bibitem{Surman+08}
R.~{Surman}, G.~C. {McLaughlin}, M.~{Ruffert}, H.-T. {Janka}, and W.~R. {Hix}.
\newblock {r-Process Nucleosynthesis in Hot Accretion Disk Flows from Black
  Hole-Neutron Star Mergers}.
\newblock {\em Astrophys.\ J.}, 679:L117--L120, June 2008.

\bibitem{Wanajo+14}
S.~{Wanajo}, Y.~{Sekiguchi}, N.~{Nishimura}, K.~{Kiuchi}, K.~{Kyutoku}, and
  M.~{Shibata}.
\newblock {Production of All the r-process Nuclides in the Dynamical Ejecta of
  Neutron Star Mergers}.
\newblock {\em Astrophys.\ J.}, 789:L39, July 2014.

\bibitem{Just+15}
O.~{Just}, A.~{Bauswein}, R.~A. {Pulpillo}, S.~{Goriely}, and H.-T. {Janka}.
\newblock {Comprehensive nucleosynthesis analysis for ejecta of compact binary
  mergers}.
\newblock {\em MNRAS}, 448:541--567, March 2015.

\bibitem{Rosswog+17}
S.~{Rosswog}, J.~{Sollerman}, U.~{Feindt}, A.~{Goobar}, O.~{Korobkin},
  C.~{Fremling}, and M.~{Kasliwal}.
\newblock {The first direct double neutron star merger detection: implications
  for cosmic nucleosynthesis}.
\newblock {\em ArXiv e-prints}, October 2017.

\bibitem{Siegel+17}
D.~M. {Siegel} and B.~D. {Metzger}.
\newblock {Three-dimensional GRMHD simulations of neutrino-cooled accretion
  disks from neutron star mergers}.
\newblock {\em ArXiv e-prints}, November 2017.

\bibitem{Tanvir+17}
N.~R. Tanvir, A.~J. Levan, C.~González-Fernández, O.~Korobkin, I.~Mandel,
  S.~Rosswog, J.~Hjorth, P.~D’Avanzo, A.~S. Fruchter, C.~L. Fryer, T.~Kangas,
  B.~Milvang-Jensen, S.~Rosetti, D.~Steeghs, R.~T. Wollaeger, Z.~Cano, C.~M.
  Copperwheat, S.~Covino, V.~D’Elia, A.~de~Ugarte~Postigo, P.~A. Evans, W.~P.
  Even, S.~Fairhurst, R.~Figuera Jaimes, C.~J. Fontes, Y.~I. Fujii, J.~P.~U.
  Fynbo, B.~P. Gompertz, J.~Greiner, G.~Hodosan, M.~J. Irwin, P.~Jakobsson,
  U.~G. Jørgensen, D.~A. Kann, J.~D. Lyman, D.~Malesani, R.~G. McMahon,
  A.~Melandri, P.~T. O’Brien, J.~P. Osborne, E.~Palazzi, D.~A. Perley,
  E.~Pian, S.~Piranomonte, M.~Rabus, E.~Rol, A.~Rowlinson, S.~Schulze,
  P.~Sutton, C.~C. Thöne, K.~Ulaczyk, D.~Watson, K.~Wiersema, and R.~A. M.~J.
  Wijers.
\newblock The emergence of a lanthanide-rich kilonova following the merger of
  two neutron stars.
\newblock {\em The Astrophysical Journal Letters}, 848(2):L27, 2017.

\bibitem{Kasen+17}
D.~{Kasen}, B.~{Metzger}, J.~{Barnes}, E.~{Quataert}, and E.~{Ramirez-Ruiz}.
\newblock {Origin of the heavy elements in binary neutron-star mergers from a
  gravitational-wave event}.
\newblock {\em \nat}, 551:80--84, November 2017.

\bibitem{Mumpower+17}
M.~R. {Mumpower}, G.~C. {McLaughlin}, R.~{Surman}, and A.~W. {Steiner}.
\newblock {Reverse engineering nuclear properties from rare earth abundances in
  the r process}.
\newblock {\em Journal of Physics G Nuclear Physics}, 44(3):034003, March 2017.

\bibitem{Beun+08}
J.~{Beun}, G.~C. {McLaughlin}, R.~{Surman}, and W.~R. {Hix}.
\newblock {Fission cycling in a supernova r process}.
\newblock {\em \prc}, 77(3):035804, March 2008.

\bibitem{Temis+15}
J.~d.~J. {Mendoza-Temis}, M.-R. {Wu}, K.~{Langanke},
  G.~{Mart{\'{\i}}nez-Pinedo}, A.~{Bauswein}, and H.-T. {Janka}.
\newblock {Nuclear robustness of the r process in neutron-star mergers}.
\newblock {\em \prc}, 92(5):055805, November 2015.

\bibitem{Sneden+08}
C.~{Sneden}, J.~J. {Cowan}, and R.~{Gallino}.
\newblock {Neutron-Capture Elements in the Early Galaxy}.
\newblock {\em Annual Review of Astronomy and Astrophysics}, 46:241--288,
  September 2008.

\bibitem{Cote+17}
B.~{C{\^o}t{\'e}}, C.~L. {Fryer}, K.~{Belczynski}, O.~{Korobkin},
  M.~{Chru{\'s}li{\'n}ska}, N.~{Vassh}, M.~R. {Mumpower}, J.~{Lippuner}, T.~M.
  {Sprouse}, R.~{Surman}, and R.~{Wollaeger}.
\newblock {The Origin of r-Process Elements in the Milky Way}.
\newblock {\em ArXiv e-prints}, October 2017.

\bibitem{Panov+03}
I.V. Panov and F.-K. Thielemann.
\newblock Final r-process yields and the influence of fission: The competition
  between neutron-induced and β-delayed fission.
\newblock {\em Nuclear Physics A}, 718(Supplement C):647 -- 649, 2003.

\bibitem{Pinedo+07}
G.~Martínez-Pinedo, D.~Mocelj, N.T. Zinner, A.~Kelić, K.~Langanke, I.~Panov,
  B.~Pfeiffer, T.~Rauscher, K.-H. Schmidt, and F.-K. Thielemann.
\newblock The role of fission in the r-process.
\newblock {\em Progress in Particle and Nuclear Physics}, 59(1):199 -- 205,
  2007.
\newblock International Workshop on Nuclear Physics 28th Course.

\bibitem{Panov+13}
I.~V. {Panov}, I.~Y. {Korneev}, G.~{Martinez-Pinedo}, and F.-K. {Thielemann}.
\newblock {Influence of spontaneous fission rates on the yields of superheavy
  elements in the r-process}.
\newblock {\em Astronomy Letters}, 39:150--160, March 2013.

\bibitem{Thielemann+83}
F.-K. {Thielemann}, J.~{Metzinger}, and H.~V. {Klapdor}.
\newblock {\em Zeitschrift fur Physik A Hadrons and Nuclei}, 309:301--317,
  December 1983.

\bibitem{Ghys+15}
L.~{Ghys}, A.~N. {Andreyev}, S.~{Antalic}, M.~{Huyse}, and P.~{Van Duppen}.
\newblock {Empirical description of {$\beta$} -delayed fission partial
  half-lives}.
\newblock {\em \prc}, 91(4):044314, April 2015.

\bibitem{Qian+02}
Y.-Z. Qian.
\newblock Neutrino-induced fission and r-process nucleosynthesis.
\newblock {\em The Astrophysical Journal Letters}, 569(2):L103, 2002.

\bibitem{Kolbe+04}
E.~Kolbe, K.~Langanke, and G.~M. Fuller.
\newblock Neutrino-induced fission of neutron-rich nuclei.
\newblock {\em Phys. Rev. Lett.}, 92:111101, Mar 2004.

\bibitem{Eichler+14}
M.~Eichler et~al.
\newblock {The Role of Fission in Neutron Star Mergers and its Impact on the
  r-Process Peaks}.
\newblock {\em Astrophys. J.}, 808(1):30, 2015.

\bibitem{Andreyev+13}
A.~N. {Andreyev}, M.~{Huyse}, and P.~{Van Duppen}.
\newblock {Colloquium: Beta-delayed fission of atomic nuclei}.
\newblock {\em Reviews of Modern Physics}, 85:1541--1559, October 2013.

\bibitem{Elseviers+13}
J.~{Elseviers}, A.~N. {Andreyev}, M.~{Huyse}, P.~{Van Duppen}, S.~{Antalic},
  A.~{Barzakh}, N.~{Bree}, T.~E. {Cocolios}, V.~F. {Comas}, J.~{Diriken},
  D.~{Fedorov}, V.~N. {Fedosseev}, S.~{Franchoo}, L.~{Ghys}, J.~A. {Heredia},
  O.~{Ivanov}, U.~{K{\"o}ster}, B.~A. {Marsh}, K.~{Nishio}, R.~D. {Page},
  N.~{Patronis}, M.~D. {Seliverstov}, I.~{Tsekhanovich}, P.~{Van den Bergh},
  J.~{Van De Walle}, M.~{Venhart}, S.~{Vermote}, M.~{Veselsk{\'y}}, and
  C.~{Wagemans}.
\newblock {{$\beta$}-delayed fission of $^{180}$Tl}.
\newblock {\em \prc}, 88(4):044321, October 2013.

\bibitem{Liberati+13}
V.~{Liberati}, A.~N. {Andreyev}, S.~{Antalic}, A.~{Barzakh}, T.~E. {Cocolios},
  J.~{Elseviers}, D.~{Fedorov}, V.~N. {Fedoseeev}, M.~{Huyse}, D.~T. {Joss},
  Z.~{Kalaninov{\'a}}, U.~{K{\"o}ster}, J.~F.~W. {Lane}, B.~{Marsh},
  D.~{Mengoni}, P.~{Molkanov}, K.~{Nishio}, R.~D. {Page}, N.~{Patronis},
  D.~{Pauwels}, D.~{Radulov}, M.~{Seliverstov}, M.~{Sj{\"o}din},
  I.~{Tsekhanovich}, P.~{Van den Bergh}, P.~{Van Duppen}, M.~{Venhart}, and
  M.~{Veselsk{\'y}}.
\newblock {{$\beta$}-delayed fission and {$\alpha$} decay of $^{178}$Tl}.
\newblock {\em \prc}, 88(4):044322, October 2013.

\bibitem{Truesdale+16}
V.~L. {Truesdale}, A.~N. {Andreyev}, L.~{Ghys}, M.~{Huyse}, P.~{Van Duppen},
  S.~{Sels}, B.~{Andel}, S.~{Antalic}, A.~{Barzakh}, L.~{Capponi}, T.~E.
  {Cocolios}, X.~{Derkx}, H.~{De Witte}, J.~{Elseviers}, D.~V. {Fedorov}, V.~N.
  {Fedosseev}, F.~P. {He{\ss}berger}, Z.~{Kalaninov{\'a}}, U.~{K{\"o}ster},
  J.~F.~W. {Lane}, V.~{Liberati}, K.~M. {Lynch}, B.~A. {Marsh}, S.~{Mitsuoka},
  Y.~{Nagame}, K.~{Nishio}, S.~{Ota}, D.~{Pauwels}, L.~{Popescu}, D.~{Radulov},
  E.~{Rapisarda}, S.~{Rothe}, K.~{Sandhu}, M.~D. {Seliverstov}, A.~M.
  {Sj{\"o}din}, C.~{Van Beveren}, P.~{Van den Bergh}, and Y.~{Wakabayashi}.
\newblock {{$\beta$} -delayed fission and {$\alpha$} decay of $^{196}$At}.
\newblock {\em \prc}, 94(3):034308, September 2016.

\bibitem{Moller+15}
P.~{M{\"o}ller}, A.~J. {Sierk}, T.~{Ichikawa}, A.~{Iwamoto}, and M.~{Mumpower}.
\newblock {\em \prc}, 91(2):024310, February 2015.

\bibitem{Mumpower+16b}
M.~R. {Mumpower}, T.~{Kawano}, and P.~{M{\"o}ller}.
\newblock {\em \prc}, 94(6):064317, December 2016.

\bibitem{Spyrou+16}
A.~{Spyrou}, S.~N. {Liddick}, F.~{Naqvi}, B.~P. {Crider}, A.~C. {Dombos}, D.~L.
  {Bleuel}, B.~A. {Brown}, A.~{Couture}, L.~{Crespo Campo}, M.~{Guttormsen},
  A.~C. {Larsen}, R.~{Lewis}, P.~{M{\"o}ller}, S.~{Mosby}, M.~R. {Mumpower},
  G.~{Perdikakis}, C.~J. {Prokop}, T.~{Renstr{\o}m}, S.~{Siem}, S.~J. {Quinn},
  and S.~{Valenta}.
\newblock {Strong Neutron-{$\gamma$} Competition above the Neutron Threshold in
  the Decay of $^{70}$Co}.
\newblock {\em Physical Review Letters}, 117(14):142701, September 2016.

\bibitem{Moller+97}
P.~{M{\"o}ller}, J.~R. {Nix}, and K.-L. {Kratz}.
\newblock {\em Atomic Data and Nuclear Data Tables}, 66:131, 1997.

\bibitem{FRDM2012a}
P.~{M{\"o}ller}, W.~D. {Myers}, H.~{Sagawa}, and S.~{Yoshida}.
\newblock {\em Physical Review Letters}, 108(5):052501, February 2012.

\bibitem{FRDM2012b}
P.~{M{\"o}ller}, A.~J. {Sierk}, T.~{Ichikawa}, and H.~{Sagawa}.
\newblock {Nuclear ground-state masses and deformations: FRDM(2012)}.
\newblock {\em Atomic Data and Nuclear Data Tables}, 109:1--204, May 2016.

\bibitem{Hill+53}
D.~L. {Hill} and J.~A. {Wheeler}.
\newblock {\em Physical Review}, 89:1102--1145, March 1953.

\bibitem{Bjornholm+73}
{\em Proc. 3rd Symp. on Physics and Chemistry of Fission}, volume~1, Rochester,
  1973. Vienna: IAEA.

\bibitem{Iljinov+92}
A.~S. {Iljinov}, M.~V. {Mebel}, N.~{Bianchi}, E.~{De Sanctis}, C.~{Guaraldo},
  V.~{Lucherini}, V.~{Muccifora}, E.~{Polli}, A.~R. {Reolon}, and P.~{Rossi}.
\newblock {Phenomenological statistical analysis of level densities, decay
  widths and lifetimes of excited nuclei}.
\newblock {\em Nuclear Physics A}, 543:517--557, July 1992.

\bibitem{Meldner+72}
H.~W. {Meldner}.
\newblock {Superheavy Element Synthesis}.
\newblock {\em Physical Review Letters}, 28:975--978, April 1972.

\bibitem{Petermann+12}
I.~{Petermann}, K.~{Langanke}, G.~{Mart{\'{\i}}nez-Pinedo}, I.~V. {Panov},
  P.-G. {Reinhard}, and F.-K. {Thielemann}.
\newblock {Have superheavy elements been produced in nature?}
\newblock {\em European Physical Journal A}, 48:122, September 2012.

\bibitem{Erler+12}
J.~{Erler}, N.~{Birge}, M.~{Kortelainen}, W.~{Nazarewicz}, E.~{Olsen}, A.~M.
  {Perhac}, and M.~{Stoitsov}.
\newblock {The limits of the nuclear landscape}.
\newblock {\em \nat}, 486:509--512, June 2012.

\bibitem{Howard+80}
W.~M. {Howard} and P.~{M{\"o}ller}.
\newblock {Calculated Fission Barriers, Ground-State Masses, and Particle
  Separation Energies for Nuclei with 76 <= Z <= 100 and 140 <= N <= 184}.
\newblock {\em Atomic Data and Nuclear Data Tables}, 25:219, 1980.

\bibitem{Jachimowicz+17}
P.~Jachimowicz, M.~Kowal, and J.~Skalski.
\newblock Adiabatic fission barriers in superheavy nuclei.
\newblock {\em Phys. Rev. C}, 95:014303, Jan 2017.

\bibitem{Mumpower+15}
M.~R. {Mumpower}, R.~{Surman}, D.-L. {Fang}, M.~{Beard}, P.~{M{\"o}ller},
  T.~{Kawano}, and A.~{Aprahamian}.
\newblock {Impact of individual nuclear masses on r -process abundances}.
\newblock {\em \prc}, 92(3):035807, September 2015.

\bibitem{Sprouse+18}
T.~{Sprouse} and M.~R. {Mumpower}.
\newblock 2018.

\bibitem{Kawano+16}
T.~{Kawano}, R.~{Capote}, S.~{Hilaire}, and P.~{Chau Huu-Tai}.
\newblock {\em \prc}, 94(1):014612, July 2016.

\bibitem{Kodama+75}
T.~{Kodama} and K.~{Takahashi}.
\newblock {\em Nuclear Physics A}, 239:489--510, March 1975.

\bibitem{NUBASE2016}
G.~Audi, F.G. Kondev, Meng Wang, W.J. Huang, and S.~Naimi.
\newblock The nubase2016 evaluation of nuclear properties.
\newblock {\em Chinese Physics C}, 41(3):030001, 2017.

\bibitem{AME2012}
G.~{Audi}, W.~{M.}, W.~{A.~H.}, K.~{F.~G.}, M.~{MacCormick}, X.~{Xu}, and
  B.~{Pfeiffer}.
\newblock {The Ame2012 atomic mass evaluation}.
\newblock {\em Chinese Physics C}, 36:2, December 2012.

\bibitem{Goriely+11}
S.~{Goriely}, A.~{Bauswein}, and H.-T. {Janka}.
\newblock {r-process Nucleosynthesis in Dynamically Ejected Matter of Neutron
  Star Mergers}.
\newblock {\em Astrophys.\ J.}, 738:L32, September 2011.

\bibitem{Boleu+72}
R.~{Boleu}, S.~G. {Nilsson}, R.~K. {Sheline}, and K.~{Takashashi}.
\newblock {On the termination of the r-process and the synthesis of superheavy
  elements}.
\newblock {\em Physics Letters B}, 40:517--521, August 1972.

\bibitem{Howard+74}
W.~M. {Howard} and J.~R. {Nix}.
\newblock {Production of Superheavy Nuclei by Multiple Capture of Neutrons}.
\newblock {\em \nat}, 247:17--20, January 1974.

\bibitem{Mumpower+12a}
M.~R. {Mumpower}, G.~C. {McLaughlin}, and R.~{Surman}.
\newblock {Formation of the rare-earth peak: Gaining insight into late-time
  r-process dynamics}.
\newblock {\em \prc}, 85(4):045801, April 2012.

\bibitem{Zagrebaev+11}
V.~I. {Zagrebaev} and W.~{Greiner}.
\newblock {\em \prc}, 83(4):044618, April 2011.

\end{thebibliography}

\end{document}